\documentclass[aps,prl,twocolumn,superscriptaddress,showpacs,floatfix]{revtex4}
\usepackage{graphicx}
\begin{document}
\title{Symmetry restrictions in chirality dependence of physical properties
of single wall nanotubes}
\author{Fei Ye}
\affiliation{Center for Advanced Study, Tsinghua University,
Beijing 100084, China}
\author{Bing-Shen Wang}
\affiliation{National Laboratory of Semiconductor Superlattice and
Microstructure
\\ and
Institute of Semiconductor, Academia Sinica, Beijing 100083,
China\\}
\author{Zhao-Bin Su}
\affiliation{Institute of Theoretical Physics, Academia Sinica, Beijing
100080, China } 
\affiliation{Center for Advanced Study, Tsinghua University, Beijing 100084,
China}
\begin{abstract}
We investigate the chirality dependence of physical properties of nanotubes
which are wrapped by the planar hexagonal lattice including graphite and
boron nitride sheet, and reveal its symmetry origin. The observables under
consideration are of scalar, vector and tensor types. These exact chirality
dependence obtained are useful to verify the experimental and numerical
results and propose accurate empirical formulas. Some important features of
physical quantities can also be extracted by only considering the symmetry
restrictions without complicated calculations.
\end{abstract}

\pacs{78.67.Ch, 61.46.+w, 73.22.-f} \maketitle
Since the discovery of carbon nanotube(CNT)\cite{Iijima}, there have been
extensive investigations on the unusual physical properties of this novel
nano-material. The simplest CNT is the single wall carbon nanotube (C-SWNT)
consisting of only one rolled up graphite sheet, which was first synthesized
in 1993\cite{Iijima2,Bethune} and now can be produced in large
scale\cite{Bronikowski, Connell}.  As an analogue to the graphite, the III-V
layered compound, namely, boron nitride(BN) sheet has also the hexagonal
lattice structure and can be wrapped into various nanotubes too. The single
wall BN nanotube(BN-SWNT) was synthesized in 1996\cite{Loiseau}, which also
attracts much attentions very recently\cite{Lauret, Ishigami}.  Unlike the
nonpolar C-SWNT which could be metallic or semiconducting with a moderate
gap\cite{Saito}, this heteropolar nanotube is expected to be always a wide
gap semiconductor\cite{Rubio,Ng}. Both of these two kinds of nanotubes have
the similar descriptions of their chiral structures and constitute promising
materials for wide applications\cite{Baughman}.

The properties of nanotubes are determined by their chiral structures.
Therefore, to investigate the chirality dependence of various physical
quantities of nanotubes is an interesting topic all the time. Particularly,
to analyze the chiral composition of bulk samples will be helpful to the
production of nanotubes with different species.  Numerous experimental and
theoretical investigations are devoted to this subject
\cite{Bachilo,Telg,Li,Wu,Ando,Kane,Reich,Yang, Ivchenko,Ye}.  Most recently,
it has been reported that detailed chirality distributions in the bulk
samples of the separated C-SWNT can be obtained by the fluorescence
measurement and by the resonant Raman spectroscopy\cite{Bachilo,Telg}. Their
assignments of the chiral number $(n_1,n_2)$ to the observed spectra are
based upon the comparison between the tight-binding (TB) calculations and
the experimental data. 

In the fluorescence experiment\cite{Bachilo}, the authors also gave the
empirical formulas of the measured von-Hove singularities with respect to
the structures of C-SWNTs, which are expressed in terms of the chiral angle
$\theta$ and chiral index $\nu$ defined as
     $\theta=\mbox{atan}\sqrt{3}n_{2}/(2n_{1}+n_{2})$, and 
	 $\nu\equiv\mbox{mod}(n_{1}-n_{2},3)$.
Actually, for the purpose to give the dependence of physical observables on
the chiral structure of nanotubes, it is more suitable and efficient  to
use $\nu$ and $\theta$ instead of the chiral numbers $(n_{1},n_{2})$.  Since
there are many types of nanotubes not being observed in the experiments or
numerical calculations, an appropriate empirical formula are quite useful
and convenient to predict the properties of those unobserved ones. Generally
speaking, there are two possible methods to obtain these empirical formulas
for different physical quantities. One is through fitting the numerical and
experimental data, the other is from the analytical expansion around the two
Dirac points of the hexagonal Brillouin zone based on various TB models and
effective mass approximations. However it has technical difficulties in the
second way to get higher order terms which are sometimes important. 

In this paper, we show the symmetry restrictions on the general chirality
dependence of physical quantities of various types such as scalar, vector
and tensor. This leads to compact forms of the chirality dependence for
these observables on $\theta$ and $\nu$. Since the results are \emph{model
independent} and exact, they not only can be used to propose accurate
empirical formulas from the numerical or experimental data, but also can
indicate some important features of physical quantities without complicated
calculations.  This idea was originated in our previous study of the natural
optical activity of C-SWNT\cite{Ye}. Our symmetry analysis is essentially
based on the hexagonal structure, and we will consider its application to
the properties of both the heteropolar BN-SWNT with large ionicity and
nonpolar C-SWNT. We examine the following physical quantities as concrete
examples to illustrate our method, such as the excitation gap, electric
polarization, dielectric tensor, and piezoelectricity. 

\begin{figure}[htpb]
    \begin{center}
        \scalebox{0.65}[0.65]{\includegraphics{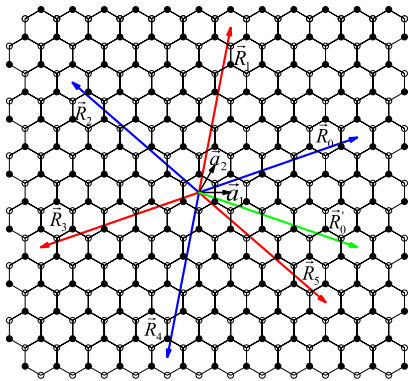}}
    \end{center}
    \caption{Illustration of hexagonal lattices. The solid and open
    circles represent boron and nitride atoms for the BN-SWNT and both
    are the carbon atoms for the C-SWNT.}
    \label{fig:sixvectors}
\end{figure}

Considering a hexagonal lattice  with base vectors $\vec{a}_{1}$ and
$\vec{a}_{2}$(see Fig.~(\ref{fig:sixvectors})), it can be rolled up into a
nanotube along the chiral vector $\vec{R}=n_{1}\vec{a}_{1} +
n_{2}\vec{a}_{2}$\cite{Saito}, so that each nanotube can be simply
represented by a pair of chiral numbers $(n_{1},n_{2})$ and the chiral angle
$\theta$ is just the angle between $\vec{R}$
and $\vec{a}_{1}$. It then can be established a mapping $f$ from the space of
chiral vectors on the planar sheet to that of the nanotube structure in a
fixed way of wrapping.
However this mapping is not one-to-one. For a given chiral vector
$\vec{R}_{0}$ with chiral angle $\theta$ on the BN sheet with $C_{3v}$
symmetry, there are another 2 equivalent vectors, obtained by rotating
$\vec{R}_{0}$ by $2\pi/3$ successively, corresponding to the same nanotube.
These vectors, as shown in Fig.~(\ref{fig:sixvectors}), have the explicit
forms
\begin{eqnarray}
    \begin{array}{l}
        \vec{R}_{0}=n_{1}\vec{a}_{1}+n_{2}\vec{a}_{2},\\
        \vec{R}_{2}=n_{2}\vec{a}_{1}-(n_{1}+n_{2})\vec{a}_{2},\\
        \vec{R}_{4}=-(n_{1}+n_2)\vec{a}_{1}+n_{1}\vec{a}_{2}
    \end{array}\;.
\end{eqnarray}
All of them constitute  an invariant subspace of the three fold symmetry of
BN sheet with the same chiral index $\nu$. Another three chiral vectors
$\vec{R}_{1}$, $\vec{R}_{3}$ and $\vec{R}_{5}$ in
Fig.~(\ref{fig:sixvectors}) have chiral angle $\theta+\pi/3$, $\theta+\pi$
and $\theta+5\pi/3$, respectively, and can be written as
\begin{eqnarray}
    \begin{array}{l}
        \vec{R}_{1}=(n_{1}+n_{2)}\vec{a}_{1}-n_{1}\vec{a}_{2},\\
        \vec{R}_{3}=-n_{1}\vec{a}_{1}-n_{2}\vec{a}_{2},\\
        \vec{R}_{5}=-n_{2}\vec{a}_{1}+(n_{1}+n_2)\vec{a}_{2}
    \end{array}\;.
\end{eqnarray}
They also form an invariant subspace, but with an opposite chiral index
$-\nu$. The nanotube mapping from $\vec{R}_{1}$ is related to that from
$\vec{R}_{0}$ by rotating the latter tube upside down. In addition, there is
another special chiral vector $\vec{R}'_{0}=(n_{1}+n_{2}) \vec{a}_{1}-n_{2}
\vec{a}_{2}$, which is the reflection of $\vec{R}_{0}$ about the base vector
$\vec{a}_1$ on the BN sheet and has the same chiral index $\nu$ of
$\vec{R}_{0}$. When mapping onto a nanotube, it corresponds to the mirror
image of that from $\vec{R}_{0}$ with respect to the section along the tube
axis.  For the graphite sheet, since the two atoms in one unit cell are the
very same, it has a higher symmetry $C_{6v}$. The six vectors $\vec{R}_{i}
(i=0,\cdots,5)$ all correspond to the same nanotube which means that a
C-SWNT may be represented by opposite chiral indices, $\nu$ and $-\nu$.

Briefly, through the mapping $f$ the manipulations on $\vec{R}$ in the
chiral vector set lead to the change of the structure in the nanotube set,
and we have the following three observations
\renewcommand{\labelenumi}{\roman{enumi}.}
\begin{enumerate}
    \item when $\theta\rightarrow\theta+{2\pi}/{3}$, the
        nanotube keeps unchanged, and $\nu$ is also unchanged.
    \item when $\theta\rightarrow\theta+{\pi}/{3}$, the
        nanotube is rotated upside down, and
    $\nu\rightarrow -\nu$.
    \item when $\theta\rightarrow-\theta$, the nanotube is
        reflected with respect to the section along the tube axis,
        and $\nu$ is unchanged.
\end{enumerate}
According to the first observation, we know that the physical quantities of
nanotube can always be expanded in terms of the triangle function of
$\theta$ with respect to each class of $\nu$
\begin{eqnarray}
    Q^{(\nu)}(\theta)=\sum_{n=0}^{\infty}a_{n}^{(\nu)}\cos(3n\theta)+b_{n}^{(\nu)}\sin(3n\theta)\;.
    \label{Fourierexpandsion}
\end{eqnarray}
$Q$ is some physical quantity. The coefficients $a_{n}^{(\nu)}$'s and
$b_{n}^{(\nu)}$'s are functions of those chirality independent variables,
such as the tube radius $r$ and some external parameters. Our analysis will
reveal \emph{ the characteristic role of  the chiral index $\nu$ in
classifying the chirality dependence}. In the following we will consider
some examples to show that the second and third observations together with
features of the physical quantity under consideration will reduce the above
chiral expressions Eq.(\ref{Fourierexpandsion}).  For the convenience of
discussion, the Cartesian coordinates for nanotube is introduced as: the
tube axis is set as $z$ direction and the cross section is the $xy$ plane
with $x$ axis passing through one atom on the tube surface.

1. \emph{Excitation Gap $\Delta$}. We can treat the band gap for the C-SWNT
and BN-SWNT as a scalar since its value does not change under rotation or
mirror reflection of the nanotube. Hence, its $\theta$-dependence must
satisfy
\begin{eqnarray} 
&&\Delta^{(\nu)}(\theta+{\pi}/{3})=\Delta^{(-\nu)}
(\theta)\nonumber\\ 
&&\Delta^{(\nu)}(-\theta)=\Delta^{(\nu)}(\theta)\;. 
\label{gapsymmetry} 
\end{eqnarray} 
Notice that in the first expression in Eq.(\ref{gapsymmetry}), $\nu$ becomes
$-\nu$ when $\theta \rightarrow \theta+\pi/3$. Then, from
Eq.(\ref{Fourierexpandsion}) the chirality dependence of $\Delta$ reads
\begin{eqnarray}
    \Delta^{(\pm)}(\theta)&=& a_{0}\pm
    a_{1}\cos(3\theta)+a_{2}\cos(6\theta)+\cdots\nonumber\\
\Delta^{(0)}(\theta)&=& a_{0}+a_{1}\cos(6\theta)+a_{2}\cos(12\theta)+\cdots\;.
    \label{gapfourier}
\end{eqnarray}
Clearly, the coefficients of $\cos(3\theta)$ for $\nu=\pm1$ should have the
same value but opposite signs.  

For the semiconducting C-SWNT with $\nu=\pm 1$, the longitudinal optical
excitation can be measured in the fluorescence experiment\cite{Bachilo} and
is quite useful in analyzing the chiral composition of bulk samples.  In
Ref.[\onlinecite{Bachilo}], the authors fitted the experimental data of the
von-Hove singularities for $\nu=1$ and $-1$ separately. According to their
fitting function, the absolute values of $a_{1}$ are quite different for
$\nu=1$ and $\nu=-1$, which does not agree with the symmetry analysis
Eq.(\ref{gapfourier}) and suggests that it may need to consider higher order
terms like $\cos(6\theta)$ for more accurate fitting functions. In fact we
have tried to fit all their data for both $\nu=\pm1$ by just one
four-parameters formula with a $\cos(6\theta)$ term
\begin{eqnarray}
\frac{p_{1}}{r}+\frac{p_{2}}{r^{2}}+\nu\frac{p_{3}\cos3\theta}{r^{2}}
+\frac{p_{4}\cos6\theta}{r^{3}}\nonumber
\end{eqnarray}
with parameters $p_i's$ to be determined, which satisfies the symmetry
restriction Eq.(\ref{gapfourier}). It turns out that the result has a
smaller root-mean-square deviation than that in Ref.  [\onlinecite
{Bachilo}].

Recently, an interesting temperature dependence of bandgap $\Delta(T)$ was
reported for the semiconducting C-SWNT in Ref.[\onlinecite{Capaz}], that is
when $\theta$ is small, the temperature dependence of gap is monotonic for
$\nu=1$ and nonmonotonic for $\nu=-1$.  The different behaviors of
$\Delta(T)$ for $\nu=\pm 1$ exist even for very close $\theta$. This could
be understood by assuming that both $a_{0}$ and $a_{1}$ are decreasing
functions of temperature. Then from Eq.(\ref{gapfourier}), it is clear that
the bandgap for $\nu=1$ is monotonically decreasing with temperature. For
$\nu=-1$ and small $\theta$, the sign of $a_{1}$ is negative, so that
$\Delta$ can be nonmonotonic as temperature varying as a result of the
interplay between $a_{0}(T)$ and $a_{1}(T)$. However when $\theta$ is close
to $\pi/6$, the $\cos(3\theta)$ is nearly vanishing and only $a_{0}$ takes
the dominant role, so that the temperature dependence of bandgap should
behave similarly for both $\nu=1$ and $-1$.

2. \emph{Electric Polarization(EP) $\vec{P}$.} The macroscopic electric
polarization along the nanotube axis is the consequence of the broken
sublattice symmetry of BN-SWNT, which was studied as a geometric phase in
Ref.[\onlinecite{Mele}]. They found the sign and size of the longitudinal
polarization are dramatically dependent on the chiral structure of nanotube.
This in fact has its symmetry origin, and we will show below that some
remarkable properties of EP can be extracted by the symmetry analysis.

Due to the helical symmetry of the nanotube\cite{ctwhite,Damnjanovic}, this
EP vector only exists in the tube axis direction($z$-axis) and vanishes in
the cross section($xy$ plane), i.e., $P_{z}\ne 0$ and $P_{x,y}=0$. According
to observations ii and iii and the vector nature of EP, we have
\begin{eqnarray}
    P^{(\nu)}_{z}(\theta+{\pi}/{3})&=&-P^{(-\nu)}_{z}(\theta)\nonumber\\
    P^{(\nu)}_{z}(-\theta)&=&P^{(\nu)}_{z}(\theta)\;,
    \label{epsymmetry}
\end{eqnarray}
which leads to the chirality dependence of $P_{z}$
\begin{eqnarray}
    P_{z}^{(0)}(\theta)&=& a_{1}\cos(3\theta)+a_{3}\cos(9\theta)+\cdots\nonumber\\
    P_{z}^{(\pm)}(\theta)&=& \pm a_{0}+ a_{1}\cos(3\theta)\pm a_{2}\cos(6\theta)+\cdots
    \label{ep}
\end{eqnarray}
Eq.(\ref{ep}) shows evidently that the armchair tubes($n_{1}=n_{2}=N$) with
$\theta=\pi/6$ and $\nu=0$ have no electric polarization. Another type of
achiral tube, zigzag tubes($n_{1}=N,n_{2}=0$), show different pictures. If
we assume reasonably the coefficients of higher order terms are small enough
comparing to the zero order term $a_{0}$, we then obtain $P^{(0)}_{z}
\approx 0$ for $\mbox{mod}(N,3)=0$, and $P^{(\pm)}_{z}\approx \pm a_{0}$ for
$\mbox{mod}(N,3)=\pm 1$, i.e., when $N$ is increasing, the EP is oscillating
among $1$, $0$ and $-1$, which is just the striking result of BN-SWNT found
in Ref.[\onlinecite{Mele}].

For the C-SWNT, the six fold $C_{6v}$ symmetry in the chiral vector set
guarantees $P^{(0)}_{z}(\theta+\pi/3) =P^{(-\nu)}_{z}(\theta)$, which
combined with Eq.~(\ref{epsymmetry}) leads to $P^{(\nu)}_{z}=0$, i.e.,
\emph{no EP in C-SWNT}.

3. \emph{Dielectric tensor $\epsilon$}. This second rank tensor is
restricted to have the following form by the helical symmetry of
nanotube\cite{Damnjanovic}:
\begin{eqnarray}
    \epsilon=\left(
    \begin{array}{ccc}
        \epsilon_{xx}&\epsilon_{xy}&0\\
        \epsilon_{yx}&\epsilon_{yy}&0\\
        0&0&\epsilon_{zz}
    \end{array}
    \right)
    \label{dtsymmtry}
\end{eqnarray}
with $\epsilon_{xx}=\epsilon_{yy}$ and $\epsilon_{xy}=-\epsilon_{yx}$. The
diagonal matrix elements of $\epsilon$, denoted by $\epsilon_{ii}$ with
$i=x,y,z$, have the same chirality dependence. Similar to the analysis of EP
vector, we obtain
\begin{eqnarray}
    &&\epsilon^{(\nu)}_{ii}(\theta+{\pi}/{3})=\epsilon^{(-\nu)}_{ii}(\theta)\nonumber\\
    &&\epsilon^{(\nu)}_{ii}(-\theta)=\epsilon^{(\nu)}_{ii}(\theta)\nonumber\\
    &&\epsilon^{(\nu)}_{xy}(\theta+{\pi}/{3})=-\epsilon^{(-\nu)}_{xy}(\theta)\nonumber\\
    &&\epsilon^{(\nu)}_{xy}(-\theta)=-\epsilon^{(\nu)}_{xy}(\theta)
    \label{dt}
\end{eqnarray}
by noticing that the diagonal term is unchanged when the tube is reversed or
reflected, and the off-diagonal term get its sign changed. Then
\begin{eqnarray}
    \epsilon_{ii}^{(\pm)}(\theta)&=&a_0\pm a_1\cos(3\theta)+a_2\cos(6\theta)+\cdots\nonumber\\
    \epsilon_{ii}^{(0)}(\theta)&=&a_0+ a_2\cos(6\theta)+a_4\cos(12\theta)+\cdots\nonumber\\
    \epsilon_{xy}^{(\pm)}(\theta)&=&b_1\sin (3\theta)\pm b_2\sin (6\theta)+\cdots\nonumber\\
    \epsilon_{xy}^{(0)}(\theta)&=&b_1\sin (3\theta)+b_3\sin(9\theta)+\cdots
    \label{dtfourier}
\end{eqnarray}
The coefficients $a_{n}'s$ and $b_{n}'s$ in the above four expressions have
no direct relationships. Obviously, the diagonal and off-diagonal terms have
quite different chirality dependence and the off-diagonal terms vanish for
the zigzag tube whose chiral angle is $0$.

The discussion above is for the heteropolar BN-SWNT. For the C-SWNT, the
higher symmetry requires $\epsilon^{(\nu)}_{xy}(\theta+{\pi}/
{3})=\epsilon^{(-\nu)}_{xy}(\theta)$, which together with Eq.~(\ref{dt})
leads to \emph{ $\epsilon^{(\nu)}_{xy}=0$ for any kind of C-SWNT}.

4. \emph{Piezoelectricity $e$.} Piezoelectricity is the response of the EP
of the material to the mechanical strain, which is a third rank tensor
defined by the derivative of EP vector with respect to the elastic strain
tensor $u$ as $e_{i,jk}={\partial P_{i}}/{\partial u_{jk}}$. 
For the quasi-one-dimensional nanotube, we are concerned with the response
of EP along the $z$ direction to uniaxial($s$) and torsional ($t$) strains,
\begin{eqnarray}
    e_{s}={\partial P_{z}}/{\partial u_{s}}, \hspace{0.5cm}
    e_{t}={\partial P_{z}}/{\partial u_{t}}\;.
    \label{piezotube}
\end{eqnarray}
$u_{s}$ is the stretch strain along the tube, and $u_{t}$ is the torsional
strain around the tube circumference. They can be related to the second rank
tensor $u_{ij}$ in the Cartesian coordinates system through 
equations $u_{s}=u_{zz}$ and $u_{t}=(xu_{zy}-yu_{zx})/{r^{2}}$
with $r^{2}=x^{2}+y^{2}$. Then, it is clear that how
$u_{s}$ and $u_{t}$ transform under the rotation of the tube upside down or
the mirror reflection with respect to the section along the tube axis.
Consequently, we have
\begin{eqnarray}
    &&e^{(\nu)}_{s}(\theta+{\pi}/{3})=-e^{(-\nu)}_{s}(\theta) \nonumber \\
    &&e^{(\nu)}_{s}(-\theta)=e^{(\nu)}_{s}(\theta) \nonumber \\
    &&e^{(\nu)}_{t}(\theta+{\pi}/{3})=-e^{(-\nu)}_{t}(\theta) \nonumber \\
    &&e^{(\nu)}_{t}(-\theta)=-e^{(\nu)}_{t}(\theta)\;,
    \label{piezosymmetry}
\end{eqnarray}
and the chirality dependence of piezoelectricity then reads
\begin{eqnarray}
    &&e^{(\pm)}_{s}(\theta)= \pm a_{0}+a_{1}\cos(3\theta)\pm a_{2}\cos(6\theta)+\cdots\nonumber\\
    &&e^{(0)}_{s}(\theta)= a_{1}\cos(3\theta)+a_{3}\cos(9\theta)+\cdots\nonumber\\
    &&e^{(\pm)}_{t}(\theta)=  b_{1}\sin(3\theta)\pm b_{2}\sin(6\theta)+b_{3}\sin(9\theta)+\cdots\nonumber\\
    &&e^{(0)}_{t}(\theta)= b_{1}\sin(3\theta)+ b_{3}\sin(9\theta)+\cdots\;.
    \label{piezoexpression}
\end{eqnarray}
Eq.~(\ref{piezoexpression}) implies the zigzag($\theta=0$) tube can only
have piezoelectric response to the longitudinal stretch and no response to
the torsion strain around the circumference. On the contrary, the armchair
($\theta=\pi/6$ and $\nu=0$) tube has response to torsion strain but not to
stretch strain. These conclusions are in consistent with the numerical
results of the \textit{ab initio} and TB calculations in
Ref.[\onlinecite{Sai}], where the chiral angle by definition has a
difference $\pi/2$ from ours. Apart from this difference in the $\theta$
definition, in Ref.[\onlinecite{Sai}] only $\cos(3\theta)$ and
$\sin(3\theta)$ terms appear in the chirality dependence of $e_{s}$ and
$e_{t}$, respectively, while there are extra terms according to our results
Eq.~(\ref{piezoexpression}). In fact, their $\theta$-dependence of $e_s$ and
$e_t$ do not agree with their numerical data very well.  The reason for this
difference is that in Ref.[\onlinecite{Sai}] the chirality dependence of
$e_{t}$ and $e_{s}$ is inherited from the BN planar sheet\cite{3theta}, so
that only the $3\theta$ terms are permitted by the $C_{3v}$ symmetry of BN
sheet. However when the sheet is rolled up, this planar $C_{3v}$ symmetry is
broken, therefore those $C_{3v}$-forbidden terms are not forbidden any more,
although they should be small by the continuous argument from sheet to
nanotube, i.e., they vanish when the tube radius tends to infinity.

For the C-SWNT, the symmetry leads to additional restriction $e^{(\nu)}
_{\alpha} (\theta+{\pi}/{3})= e^{(-\nu)} _{\alpha} (\theta)$, which combined
with Eq.~(\ref{piezosymmetry}) requires $e^{(\nu)}_{\alpha} (\theta)=0$ for
both $\alpha=s$ and $t$, namely, \emph{for C-SWNT there should be no
piezoelectricity due to its nonpolar feature.}

In the above discussion, the expansion coefficients $a^{(\nu)}_{n}$'s and
$b^{(\nu)}_{n}$'s can not be determined by the symmetry argument which in
fact depend on the chirality independent parameters, such as the tube
diameter, magnetic flux, temperature, and so on. By tuning the external
parameters, one can adjust the magnitude of $a^{(\nu)}_{n}$'s and
$b^{(\nu)}_{n}$'s, which will be helpful to identify the chirality of the
tubes. As examples, one could refer to Ref.[\onlinecite{Ye,Capaz}] to find the
different dependence of these coefficients on magnetic flux or temperature.

As a conclusion, we give the explicit $\theta$-dependence of physical
quantities for different values of $\nu$ by a symmetry analysis. It shows
clearly that the chiral index $\nu$ plays a characteristic role in
describing the chirality dependence. This model independent method may be
used to verify the numerical and experimental data and also can give rise to
some important properties qualitatively without complicated calculations.
In addition, this method is not restricted to the examples illustrated in
this paper and could be extended to other situations.

The author F. Ye would like to thank Dr. J. Wu at CASTU for helpful
discussion.


\begin{thebibliography}{}
	\bibitem{Iijima} S. Iijima, Nature \textbf{354}, 56(1991)
	\bibitem{Iijima2} S. Iijima and T. Ichihashi, Nature \textbf{363},
	603(1993)
	\bibitem{Bethune} D. S. Bethune, \textit{et al.}, Nature \textbf{363},
	605(1993)
    \bibitem{Bronikowski} M. J. Bronikowski, \textit{et al.}, J. Vac. Sci. Technol. A \textbf{19}, 1800(2001)
    \bibitem{Connell} M. O' Connell \textit{et al.}, Science \textbf{297}, 593(2002)
    \bibitem{Loiseau} A. Loiseau, F. Willaime, N. Demoncy, G. Hug, and H.
	Pascard, Phys. Rev. Lett.
	\textbf{76}, 4737(1996)
    \bibitem{Lauret} J. S. Lauret, \textit{et. al.}, Phys. Rev. Lett.
    \textbf{94}, 037405(2005)
    \bibitem{Ishigami} M. Ishigami, J. D. Sau, S. Aloni, M. L. Cohen, and A.
	Zettl, Phys. Rev. Lett.
    \textbf{94}, 056804(2005)
    \bibitem{Saito} S. Saito, G.Dresselhaus and M. S. Dresselhaus, {\it Physical Properties of Carbon Nanotube}(Imperial College Press, London, 1998)
    \bibitem{Rubio} A. Rubio, J. L. Corkill, and M. L. Cohen, Phys. Rev. B
	\textbf{49}, R5081(1994)
    \bibitem{Ng} M. F. Ng and R. Q. Zhang, Phys. Rev. B \textbf{69},
    115417(2004)
	\bibitem{Baughman} For the potential applications of CNT, one can refer
	to the review article, R. H. Baughman, A. A. Zakhidov, and W. A. de Heer,
	Science \textbf{297}, 787(2002)
	\bibitem{Bachilo} Sergei M. Bachilo, \textit{et al.}  Science \textbf{298}, 2361(2002)
    \bibitem{Telg}H. Telg, J. Maultzsch, S. Reich, F. Hennrich, and C. Thomsen,
         Phys. Rev. Lett. \textbf{93}, 177401 (2004)
    \bibitem{Li} Lain-Jong Li, R. J. Nicholas, R. S Deacon, and P. A. Shields, Phys. Rev. Lett. \textbf{93}, 156104(2004)
    \bibitem{Wu} J. Wu, \textit{et al}, Phys. Rev. Lett. \textbf{93}, 017404(2004)
    \bibitem{Ando} H. Ajiki and T. Ando, J. Phys. Soc. Jap. \textbf{65}, 505(1996)
    \bibitem{Kane} C. L. Kane and E. J. Mele, Phys. Rev. Lett. \textbf{78}, 1932(1997)
    \bibitem{Reich} S. Reich and C. Thomsen, Phys. Rev. B \textbf{62}, 4273(2000)
    \bibitem{Yang} Liu Yang and Jie Han, Phys. Rev. Lett. \textbf{85}, 154(2000)
    \bibitem{Ivchenko} E. L. Ivchenko and B. Spivak, Phys. Rev. B \textbf{66}, 155404(2002)
    \bibitem{Ye} Fei Ye, Bing-Shen Wang and Zhao-Bin Su, Phys. Rev. B \textbf{70}, 153406(2004)
    \bibitem{Capaz} Rodrigo B. Capaz, C. D. Spataru, P. Tangney, M. L.
	Cohen, and S. G. Louie, Phys. Rev. Lett.
    \textbf{94}, 036801(2005)
    \bibitem{Sai} Na Sai and E.J. Mele, Phys.  Rev. B \textbf{68}, 241405(R)(2003)
    \bibitem{Mele} E. J. Mele and Petr Kr\'al,  Phys. Rev. Lett. \textbf{88}, 056803(2002)
    \bibitem{ctwhite} C. T. White, D. H. Robertson and J. W. Mintmire, Phys. Rev. B \textbf{47}, R5485(1993)
    \bibitem{Damnjanovic}M. Damnjanovi\'{c}, I. Milo\v{s}evi\'c, T. Vukovi\'c and R. Sredanovi\'c, Phys. Rev. B \textbf{60}, 2728(1999)
	\bibitem{3theta} It can be obtained through a rotation by $\theta$ of the piezoelectric tensor of the BN sheet.
\end{thebibliography}
\end{document}